\documentclass[aps,prd,preprint,preprintnumbers,superscriptaddress,showpacs]{revtex4-1}
\usepackage{graphicx}

\usepackage{ulem}
\usepackage{color}
\definecolor{My_red}        {cmyk}{0.00,1.00,1.00,0.20}

\newcommand{\bmat}{\left(\begin{array}}
\newcommand{\emat}{\end{array}\right)}
\newcommand{\beq}{\begin{equation}}
\newcommand{\eeq}{\end{equation}}

\def\bwt{\begin{widetext}}
\def\ewt{\end{widetext}}
\def\be{\begin{equation}}
\def\ee{\end{equation}}
\def\bea{\begin{eqnarray}}
\def\eea{\end{eqnarray}}
\def\bean{\begin{eqnarray*}}
\def\eean{\end{eqnarray*}}
\def\bary{\begin{array}}
\def\eary{\end{array}}
\def\bit{\begin{itemize}}
\def\eit{\end{itemize}}

\def\su5u1{SU(5) \times U(1)}
\def\fsu5u1{SU(5) \times U(1)'}
\def\so10{SO(10)}
\def\sq20{SO(10) \times SO(10)}

\def\bwt{\begin{widetext}}
\def\ewt{\end{widetext}}
\def\be{\begin{equation}}
\def\ee{\end{equation}}
\def\bea{\begin{eqnarray}}
\def\eea{\end{eqnarray}}
\def\bean{\begin{eqnarray*}}
\def\eean{\end{eqnarray*}}
\def\bary{\begin{array}}
\def\eary{\end{array}}
\def\bit{\begin{itemize}}
\def\eit{\end{itemize}}

\def\su5u1{SU(5) \times U(1)}
\def\fsu5u1{SU(5) \times U(1)'}
\def\so10{SO(10)}
\def\sq20{SO(10) \times SO(10)}

\usepackage[centertags]{amsmath}
\usepackage{amssymb}

\begin{document}

\title{A Flippon Related Singlet at the LHC II}

\author{Tianjun Li}

\affiliation{State Key Laboratory of Theoretical Physics and Kavli Institute for Theoretical Physics China (KITPC), Institute of Theoretical Physics, Chinese Academy of Sciences, Beijing 100190, P. R. China}

\affiliation{
  School of Physical Sciences, University of Chinese Academy of Sciences,
  Beijing 100049, P. R. China
}

\affiliation{
School of Physical Electronics, University of Electronic Science and Technology of China, Chengdu 610054, P. R. China}

\author{James A. Maxin}

\affiliation{Department of Physics and Engineering Physics, The University of Tulsa, Tulsa, OK 74104, USA}

\author{Van E. Mayes}

\affiliation{Department of Physics, University of Houston-Clear Lake, Houston, TX 77058, USA}

\author{Dimitri V. Nanopoulos}

\affiliation{George P. and Cynthia W. Mitchell Institute for Fundamental Physics and Astronomy, Texas A$\&$M University, College Station, TX 77843, USA}

\affiliation{Astroparticle Physics Group, Houston Advanced Research Center (HARC), Mitchell Campus, Woodlands, TX 77381, USA}

\affiliation{Academy of Athens, Division of Natural Sciences, 28 Panepistimiou Avenue, Athens 10679, Greece}

\date{\today}

\begin{abstract}

We consider the 750~GeV diphoton resonance at the 13~TeV LHC in the ${\cal F}$-$SU(5)$ model with a Standard Model (SM) singlet field which couples to TeV-scale vector-like particles, dubbed $flippons$. This singlet field assumes the role of the 750~GeV resonance, with production
via gluon fusion and subsequent decay to a diphoton via the vector-like particle loops. We present a numerical analysis showing that the observed 8~TeV and 13~TeV diphoton production cross-sections can be generated in the model space with realistic electric charges and Yukawa couplings for light vector-like masses. We further discuss the experimental viability of light vector-like masses in a General No-Scale ${\cal F}$-$SU(5)$ model, offering a few benchmark scenarios in this consistent GUT that can satisfy all experimental constraints imposed by the LHC and other essential experiments.

\end{abstract}


\preprint{ACT-01-16}

\maketitle

\section{Introduction}

The ATLAS~\cite{bib:ATLAS_diphoton} and CMS~\cite{bib:CMS_diphoton} Collaborations recently announced an excess of events in the diphoton channel with invariant mass of about 750~GeV at the 13~TeV LHC II. Assuming a narrow width resonance with an integrated luminosity of 3.2 ${\rm fb}^{-1}$, the ATLAS Collaboration reported observations of a local $3.6\sigma$ excess at a diphoton invariant mass of around 747~GeV. The signal significance elevates to $3.9\sigma$ with a preferred width of about 45~GeV when considering a wider width resonance. In parallel, the CMS Collaboration also observed a diphoton excess with a local significance of $2.6\sigma$ at an invariant mass of around 760~GeV at an integrated luminosity of 2.6 ${\rm fb}^{-1}$, however, the CMS signal significance reduces to $2\sigma$ when assuming a decay width of around 45~GeV.
The corresponding diphoton excesses in the production cross sections can be roughly estimated as
$\sigma_{pp\to \gamma \gamma}^{13~ {\rm TeV}} \sim 3-13~{\rm fb}$~~\cite{bib:ATLAS_diphoton, bib:CMS_diphoton}. Moreover, the CMS Collaboration completed a similar search for diphoton resonances~\cite{Khachatryan:2015qba} at the $\sqrt{s} =$ 8~TeV LHC I, observing a slight excess $\sim$ 2$\sigma$ at an invariant mass of about 750~GeV. Nonetheless, ATLAS did not probe beyond the mass of 600~GeV for this channel~\cite{Aad:2014ioa}. Therefore, the current ATLAS and CMS results at $\sqrt{s} =$~13~TeV are indeed consistent with those at $\sqrt{s} =$~8~TeV in the diphoton channel. Although the excess presently remains below the statistically significant threshold, it has drawn substantial attention from the particle physics community, resulting in a multitude of diverse explanations ranging from singlets, axions, and extended Higgs sectors to dark matter, etc~\cite{Dutta:2015wqh,Falkowski,diphoton-first,Franceschini:2015kwy,diphoton-rest,Hall:2015xds,Patel:2015ulo,Ding:2015rxx,Allanach:2015ixl, diphoton-New}. 

Our work here will focus on a supersymmetric description of the diphoton resonance. Supersymmetry (SUSY) provides an elegant solution to the gauge hierarchy problem in the Standard Model (SM), offering numerous appealing features. In Supersymmetric SMs (SSMs) with R-parity, gauge coupling unification can be realized, the electroweak gauge symmetry can be broken radiatively, the Lightest Supersymmetric Particle (LSP), namely the neutralino, provides a suitable dark matter candidate, etc. Of particular significance, gauge coupling unification strongly supports Grand Unified Theories (GUTs), where supersymmetry can bridge low energy phenomenology to high-energy fundamental physics.

String-scale gauge coupling unification was achieved through our proposed testable flipped $SU(5)\times U(1)_X$ models~\cite{smbarr, dimitri, AEHN-0} with TeV-scale vector-like particles~\cite{Jiang:2006hf}, dubbed flippons, which were then subsequently constructed from local F-theory model building~\cite{Jiang:2008yf, Jiang:2009za}. These types of models can be realized in free-fermionic string constructions too~\cite{LNY}, referred to as ${\cal F}$-$SU(5)$. A brief review of the ``miracles'' of flippons in ${\cal F}$-$SU(5)$ is in order. First, the lightest CP-even Higgs boson mass can be lifted to 125~GeV rather easily due to the one-loop contributions from the Yukawa couplings between the vector-like particles (flippons) and Higgs fields~\cite{Huo:2011zt, Li:2011ab}. Second, although the dimension-five proton decays mediated by colored Higgsinos are highly suppressed due to the missing partner mechanism and TeV-scale $\mu$ term, the dimension-six proton decays via the heavy gauge boson exchanges are within the reach of the future proton decay experiments such as the Hyper-Kamiokande experiment. This is due to the fact that the $SU(3)_C\times SU(2)_L$ gauge couplings are still unified at the traditional GUT scale while the unified gauge couplings become larger due to vector-like particle contributions~\cite{Li:2009fq, Li:2010dp}. This is in large contrast to the minimal flipped $SU(5)\times U(1)_X$ model, whose proton lifetime is too lengthy for future proton decay experiments. Third, we can consider No-Scale supergravity~\cite{Cremmer:1983bf} as a result of the string unification scale. Specifically, the lightest neutralino fulfills the role of the LSP and is lighter than the light stau due to the longer running of the Renormalization Group Equations (RGEs), providing the LSP as a dark matter candidate~\cite{Li:2010ws, Li:2010mi, Li:2011xua}. Fourth, given No-Scale supergravity, there exists a distinctive mass ordering $M({\tilde t}_1) < M({\tilde g}) < M({\tilde q}) $ of a light stop and gluino, with both substantially lighter than all other squarks~\cite{Li:2010ws, Li:2010mi, Li:2011xua}. A primary consequence of this SUSY spectrum mass pattern at the LHC is the prediction of large multijets events~\cite{Li:2011hya}. Fifth, with a merging of both No-Scale supergravity and the Giudice-Masiero (GM) mechanism~\cite{Giudice:1988yz}, the supersymmetry electroweak fine-tuning problem can be elegantly solved rather naturally~\cite{Leggett:2014mza, Leggett:2014hha}. In this paper, we shall demonstrate another ``miracle'' of the flippons: With the addition of a SM singlet field $S$ with mass about 750~GeV, we can explain the 750~GeV diphoton excess. The deep fundamental point we wish to emphasize here is that
with flippons in the loops, the singlet $S$ can be produced via gluon fusion~\cite{Georgi:1977gs}
and then consequently decay into a diphoton pair~\cite{Ellis:1975ap}.

\section{Brief Review of ${\cal F}$-$SU(5)$ Models }

First we shall briefly review the minimal flipped $SU(5)$ model~\cite{smbarr, dimitri, AEHN-0}. The gauge group for the flipped $SU(5)$ model is
$SU(5)\times U(1)_{X}$, which can be embedded into the $SO(10)$ model. We define the generator $U(1)_{Y'}$ in $SU(5)$ as 
\bea 
T_{\rm U(1)_{Y'}}={\rm diag} \left(-\frac{1}{3}, -\frac{1}{3}, -\frac{1}{3},
 \frac{1}{2},  \frac{1}{2} \right).
\label{u1yp}
\eea
and the hypercharge is given by
\bea
Q_{Y} = \frac{1}{5} \left( Q_{X}-Q_{Y'} \right).
\label{ycharge}
\eea
There are three families of the SM fermions whose quantum numbers under $SU(5)\times U(1)_{X}$ are respectively
\bea
F_i={\mathbf{(10, 1)}},~ {\bar f}_i={\mathbf{(\bar 5, -3)}},~
{\bar l}_i={\mathbf{(1, 5)}},
\label{smfermions}
\eea
where $i=1, 2, 3$. The SM particle assignments in $F_i$, ${\bar f}_i$ and ${\bar l}_i$ are
\bea
F_i=(Q_i, D^c_i, N^c_i),~{\overline f}_i=(U^c_i, L_i),~{\overline l}_i=E^c_i~,~
\label{smparticles}
\eea
where $Q_i$ and $L_i$ are respectively the superfields of the left-handed quark and lepton doublets, $U^c_i$, $D^c_i$, $E^c_i$ and $N^c_i$ are the $CP$ conjugated superfields for the right-handed up-type quarks, down-type quarks, leptons and neutrinos, respectively. To generate the heavy right-handed neutrino masses, we can introduce three SM singlets $\phi_i$.

The breaking of the GUT and electroweak gauge symmetries results from introduction of two pairs of Higgs representations
\bea
H={\mathbf{(10, 1)}},~{\overline{H}}={\mathbf{({\overline{10}}, -1)}},
~h={\mathbf{(5, -2)}},~{\overline h}={\mathbf{({\bar {5}}, 2)}}.
\label{Higgse1}
\eea
We label the states in the $H$ multiplet by the same symbols as in the $F$ multiplet, and for ${\overline H}$ we just add ``bar'' above the fields.
Explicitly, the Higgs particles are
\bea
H=(Q_H, D_H^c, N_H^c)~,~
{\overline{H}}= ({\overline{Q}}_{\overline{H}}, {\overline{D}}^c_{\overline{H}}, 
{\overline {N}}^c_{\overline H})~,~\,
\label{Higgse2}
\eea
\bea
h=(D_h, D_h, D_h, H_d)~,~
{\overline h}=({\overline {D}}_{\overline h}, {\overline {D}}_{\overline h},
{\overline {D}}_{\overline h}, H_u)~,~\,
\label{Higgse3}
\eea
where $H_d$ and $H_u$ are one pair of Higgs doublets in the MSSM. We also add one SM singlet $\Phi$.

The $SU(5)\times U(1)_{X}$ gauge symmetry is broken down to the SM gauge symmetry by introduction of the following Higgs superpotential at the GUT scale
\bea
{\it W}_{\rm GUT}=\lambda_1 H H h + \lambda_2 {\overline H} {\overline H} {\overline
h} + \Phi ({\overline H} H-M_{\rm H}^2)~.~ 
\label{spgut} 
\eea
There is only one F-flat and D-flat direction, which can always be rotated along the $N^c_H$ and ${\overline {N}}^c_{\overline H}$ directions. Therefore, we obtain $<N^c_H>=<{\overline {N}}^c_{\overline H}>=M_{\rm H}$. In addition, the superfields $H$ and ${\overline H}$ are eaten and acquire large masses via the supersymmetric Higgs mechanism, except for $D_H^c$ and ${\overline {D}}^c_{\overline H}$. Furthermore, the superpotential terms $ \lambda_1 H H h$ and $ \lambda_2 {\overline H} {\overline H} {\overline h}$ couple the $D_H^c$ and
${\overline {D}}^c_{\overline H}$ with the $D_h$ and ${\overline {D}}_{\overline h}$, respectively, to form the massive eigenstates with masses
$2 \lambda_1 <N_H^c>$ and $2 \lambda_2 <{\overline {N}}^c_{\overline H}>$. As a consequence, we naturally have the doublet-triplet splitting due to the missing partner mechanism~\cite{AEHN-0}. The triplets in $h$ and ${\overline h}$ only have
small mixing through the $\mu$ term, hence, the Higgsino-exchange mediated proton decay is negligible, {\it i.e.},
there is no dimension-5 proton decay problem. 

String-scale gauge coupling unification~\cite{Jiang:2006hf, Jiang:2008yf, Jiang:2009za} is achieved by the introduction of the following vector-like particles (flippons) at the TeV scale
\begin{eqnarray}
&& XF ={\mathbf{(10, 1)}}~,~{\overline{XF}}={\mathbf{({\overline{10}}, -1)}}~,~\\
&& Xl={\mathbf{(1, -5)}}~,~{\overline{Xl}}={\mathbf{(1, 5)}}~.~\,
\end{eqnarray}
The particle content from the decompositions of $XF$, ${\overline{XF}}$, $Xl$, and ${\overline{Xl}}$ under the SM gauge symmetry are
\begin{eqnarray}
&& XF = (XQ, XD^c, XN^c)~,~ {\overline{XF}}=(XQ^c, XD, XN)~,~\\
&& Xl= XE~,~ {\overline{Xl}}= XE^c~.~
\end{eqnarray}
Under the $SU(3)_C \times SU(2)_L \times U(1)_Y$ gauge symmetry, the quantum numbers for the extra vector-like particles are
\begin{eqnarray}
&& XQ={\mathbf{(3, 2, \frac{1}{6})}}~,~
XQ^c={\mathbf{({\bar 3}, 2,-\frac{1}{6})}} ~,~\\
&& XD={\mathbf{({3},1, -\frac{1}{3})}}~,~
XD^c={\mathbf{({\bar 3},  1, \frac{1}{3})}}~,~\\
&& XN={\mathbf{({1},  1, {0})}}~,~
XN^c={\mathbf{({1},  1, {0})}} ~,~\\
&& XE={\mathbf{({1},  1, {-1})}}~,~
XE^c={\mathbf{({1},  1, {1})}}~.~\,
\label{qnum}
\end{eqnarray}

In the ${\cal F}$-$SU(5)$ from the free-fermionic string constructions~\cite{LNY} and
local F-theory model building~\cite{Jiang:2008yf, Jiang:2009za, Heckman:2008es}, we can have the SM singlet
fields, which arise from the two seven-brane intersections and can couple to the vector-like particles. Thus,
to explain the 750~GeV diphoton excess, we introduce a SM singlet $S$ with mass
about $M_S \approx 750$~GeV. The superpotential for the flippons and $S$ is
\begin{eqnarray}
  W&=& \lambda_Q S XQ XQ^c + \lambda_D S XD XD^c + \lambda_N S XN XN^c  +\lambda_E S XE XE^c
+ \lambda_i^Q S Q_i XQ^c
\nonumber \\ &&
+ \lambda_i^D S XD D_i^c 
  +\lambda_i^E S XE E_i^c + y_D XQ XD^c H_d + y'_D XQ^c XD H_u
  +y_i^Q Q_i XD^c H_d
\nonumber \\ &&
  + y_i^D XQ D_i^c H_d + y_i^L L_i XE^c H_d
  + y_i^N L_i XN^c H_u + \lambda S H_d H_u
  + M_{XQ} XQ XQ^c
\nonumber \\ &&
+ M_{XD} XD XD^c + M_{XN} XN XN^c + M_{XE} XE XE^c + \mu_i^Q  Q_i XQ^c + \mu_i^D  XD D_i^c
\nonumber \\ &&
  + \mu_i^E XE E_i^c~.~
\end{eqnarray}
From above superpotential, we can show that the flippons can mix with the SM fermions and
decay to the SM particles and Higgs fields. In particular, to avoid the mixings between
$S$ and $H_d/H_u$, we assume that $\lambda$ is small, which can be realized by adjusting
the seven-brane configuration properly in the local F-theory model building.

The relevant supersymmetry breaking soft terms for diphoton excesses are
\begin{eqnarray}
  V_{\rm soft} &=& {\widetilde M}^2_{XQ} (|{\widetilde{XQ}}|^2+ |{\widetilde{XQ}^c}|^2)
  + {\widetilde M}^2_{XD} (|{\widetilde{XD}}|^2+ |{\widetilde{XD}^c}|^2)
  + {\widetilde M}^2_{XN} (|{\widetilde{XN}}|^2+ |{\widetilde{XN}^c}|^2)
  \nonumber \\
  && + {\widetilde M}^2_{XE} (|{\widetilde{XE}}|^2+ |{\widetilde{XE}^c}|^2)
  -\left(\lambda_Q A_Q S \widetilde{XQ} {\widetilde{XQ}^c}
  +\lambda_D A_D S \widetilde{XD} {\widetilde{XD}^c}
  \right. \nonumber \\   && \left.
+\lambda_N A_N S \widetilde{XN} {\widetilde{XN}^c}
+\lambda_E A_E S \widetilde{XE} {\widetilde{XE}^c}
+B_{XQ}  M_{XQ} XQ XQ^c
 \right. \nonumber \\   && \left.
+ B_{XD} M_{XD} XD XD^c + B_{XN} M_{XN} XN XN^c
+ B_{XE} M_{XE} XE XE^c +{\rm H.C.}\right)~.~\,
\end{eqnarray}

\section{Diphotons at the $\sqrt{s} = 13$~TeV LHC}

The production cross-sections of a 750~GeV resonance observed by CMS are $\sigma(pp \rightarrow S \rightarrow \gamma \gamma) = 0.5 \pm 0.6~{\rm fb}$ at $\sqrt{s} = 8$~TeV~\cite{Khachatryan:2015qba}and $6 \pm 3~{\rm fb}$ at $\sqrt{s} = 13$~TeV~\cite{bib:ATLAS_diphoton}, in conjunction with the ATLAS observations of $\sigma(pp \rightarrow S \rightarrow \gamma \gamma) = 0.4 \pm 0.8~{\rm fb}$ at $\sqrt{s} = 8$~TeV~\cite{Aad:2014ioa} and $10 \pm 3~{\rm fb}$ at $\sqrt{s} = 13$~TeV~\cite{bib:ATLAS_diphoton}. A diphoton invariant mass of $M_S \approx 750$~GeV indicates a best-fit decay width of about $\Gamma \sim 45$~GeV, though in this work we shall adopt a less restrictive scenario, constraining the total decay width to $\Gamma \sim 5 - 45$~GeV. For a spin-0 resonance, the observed cross-section can be accounted for with a 45~GeV decay width if~\cite{Franceschini:2015kwy}
\begin{equation}
\frac{\Gamma_{\gamma \gamma} \Gamma_{gg}}{M_S^2} \approx 1.1 \times 10^{-6} \frac{\Gamma}{M_S} \approx 6 \times 10^{-8}
\label{gg_M2_45}
\end{equation}
and for a 5~GeV decay width if
\begin{equation}
\frac{\Gamma_{\gamma \gamma} \Gamma_{gg}}{M_S^2} \approx 7 \times 10^{-9}
\label{gg_M2_5}
\end{equation}
where we employ the compact notation $\Gamma_{\gamma \gamma} = \Gamma (S \rightarrow \gamma \gamma)$ and $\Gamma_{gg} = \Gamma (S \rightarrow gg)$. These conditions shall be implemented to constrain the model to achieve the observed production cross-sections, though in our calculations we shall relax the Eq. (\ref{gg_M2_45}) and Eq. (\ref{gg_M2_5}) constraints to $\gtrsim 10^{-9}$. 

The effective loop-level couplings amongst the Standard Model gauge bosons and scalar $S$ are given by
\begin{equation}
-{\cal L} = \frac{S}{M_S} \left[ \kappa_{EM} F_{\mu \nu}^{EM} F^{EM \smallskip \mu \nu} + \kappa_3 G_{\mu \nu}^a G^{\mu \nu \, a}   \right]
\label{lagrangian}
\end{equation}
where $F_{\mu \nu}^{EM}$ and $G_{\mu \nu}^a$ are the photon and gluon field strength tensors, respectively, with $a = 1,~2,..8$. The effective operators are represented by $\kappa_{EM}$ and  $\kappa_3$, which are written as
\begin{equation}
\kappa_{EM} = \frac{\alpha_{EM}}{4 \pi} \left[ \sum_{f} \frac{\lambda_f M_S}{M_f} Q_f^2 N_{EM}^f F_f + \sum_{\widetilde{f}} \frac{\lambda_f A_f M_S}{M_{\widetilde{f}}^2} Q_{\widetilde{f}}^2 N_{EM}^f F_{\widetilde{f}} \right]
\label{kem}
\end{equation}
\begin{equation}
\kappa_{3} = \frac{\alpha_{3}}{4 \pi} \left[ \sum_{f} \frac{\lambda_f M_S}{M_f} N_{3}^f F_f + \sum_{\widetilde{f}} \frac{\lambda_f A_f M_S}{M_{\widetilde{f}}^2} N_{3}^f F_{\widetilde{f}} \right]
\label{k3}
\end{equation}
where $N_{EM}^f$ is the $SU(3)_C$ color factor, $N_{3}^f$ the $SU(2)_L$ doublet factor, $Q_U = 2/3$, $Q_D = -1/3$, $Q_E = -1$, and the functions $F_f$ and $F_{\widetilde{f}}$ are expressed as
\begin{eqnarray}
&& F_f = 2 \chi \left[ 1 + ( 1- \chi) f(\chi) \right] \\
&& F_{\widetilde{f}} = \chi \left[ -1 + \chi f(\chi) \right] \
\label{feq}
\end{eqnarray}
with the function $\chi$ denoted by
\begin{equation}
\chi = 4  \frac{M_{f/ \widetilde{f}}^2}{M_S^2}
\label{chi}
\end{equation}
The triangle loop functions $f(\chi)$ are defined here as
\begin{eqnarray}
f(\chi)=
	\begin{cases}
		\arcsin^2[\sqrt{\chi^{-1}}]  & \mbox{if } \chi \geq 1 \\
		-\frac{1}{4} \left[ \ln \frac{1 + \sqrt{1 - \chi}}{1 - \sqrt{1 - \chi}} - i \pi \right]^2 & \mbox{if } \chi < 1.
	\end{cases}
\label{loop_function}
\end{eqnarray}
The diphoton and digluon decay widths in ${\cal F}$-$SU(5)$ are computed from
\begin{equation}
\Gamma_{\gamma \gamma} = \frac{\left| \kappa_{EM} \right|^2}{4 \pi} M_S
\label{Gamma_pp}
\end{equation}
\begin{equation}
\Gamma_{gg} = \frac{2 \left| \kappa_{3} \right|^2}{\pi} M_S
\label{Gamma_gg}
\end{equation}

\noindent The diphoton production cross-section is calculated from
\begin{equation}
\sigma(pp \rightarrow S \rightarrow \gamma \gamma) = \frac{K C_{gg} \Gamma(S \rightarrow gg) \Gamma(S \rightarrow \gamma \gamma)}{s \Gamma  M_S}
\label{xsection_pp}
\end{equation}

\begin{table}[htbp!]
\centering
\tiny
\caption{Decay widths and production cross-sections for a total decay width of $\Gamma = 5$~GeV for some sample points. All masses and decay widths are in GeV. The cross-sections are in femtobarns (fb). The ${\rm Br}_{\rm DM}$ represents the branching ratio allocated to dark matter. For simplicity, we assume here that $M_{XE} = M_{XN}$ and $\lambda_E = \lambda_N$.}
\begin{tabular}{|c|c|c|c|c|c|c|c|c|c|c|c||c|c|c||c|c|c|c||c|c|}\cline{1-21}
\multicolumn{21}{|c|}{$\Gamma = 5~ {\rm GeV}$} \\ \hline
$M_{XU}$&$M_{XD}$&$M_{XE}$&$M_{\widetilde{XU}}$&$M_{\widetilde{XD}}$&$M_{\widetilde{XE}}$&$\lambda_{U}$&$\lambda_{D}$&$\lambda_E$&$A_{U}$&$A_{D}$&$A_E$&$\Gamma_{\gamma \gamma}$&$\Gamma_{gg}$&$\Gamma_{XE+XN}$&${\rm Br}_{\gamma \gamma}$&${\rm Br}_{gg}$&${\rm Br}_{XE+XN}$&${\rm Br}_{\rm DM}$&$\sigma_{\gamma \gamma}^{8~{\rm TeV}}$&$\sigma_{\gamma \gamma}^{13~{\rm TeV}}$\\ \hline \hline
$	900	$&$	800	$&$	230	$&$	1273	$&$	1131	$&$	325	$&$	0.83	$&$	0.70	$&$	0.31	$&$	2700	$&$	2400	$&$	1380	$&$	0.0011	$&$	0.64	$&$	1.41	$&$	0.00023	$&$	0.13	$&$	0.28	$&$	0.59	$&$	0.41	$&$	1.89	$	\\ \hline
$	860	$&$	740	$&$	260	$&$	1216	$&$	1047	$&$	368	$&$	0.83	$&$	0.80	$&$	0.31	$&$	2500	$&$	2200	$&$	1560	$&$	0.0016	$&$	0.88	$&$	1.07	$&$	0.00032	$&$	0.18	$&$	0.21	$&$	0.61	$&$	0.79	$&$	3.69	$	\\ \hline
$	1000	$&$	1000	$&$	265	$&$	1414	$&$	1414	$&$	375	$&$	0.83	$&$	0.83	$&$	0.31	$&$	3000	$&$	3000	$&$	1590	$&$	0.0014	$&$	0.53	$&$	1.02	$&$	0.00029	$&$	0.11	$&$	0.20	$&$	0.69	$&$	0.43	$&$	2.01	$	\\ \hline
$	900	$&$	1000	$&$	265	$&$	1273	$&$	1414	$&$	375	$&$	0.80	$&$	0.83	$&$	0.31	$&$	2700	$&$	3000	$&$	1590	$&$	0.0015	$&$	0.56	$&$	1.02	$&$	0.00030	$&$	0.11	$&$	0.20	$&$	0.68	$&$	0.47	$&$	2.18	$	\\ \hline
$	860	$&$	740	$&$	400	$&$	1216	$&$	1047	$&$	566	$&$	0.83	$&$	0.83	$&$	0.31	$&$	2500	$&$	2200	$&$	2400	$&$	0.0005	$&$	0.93	$&$	0.00	$&$	0.00010	$&$	0.19	$&$	0.00	$&$	0.81	$&$	0.27	$&$	1.24	$	\\ \hline
\end{tabular}
\label{tab:5GeV}
\end{table}

\begin{table}[htbp!]
\centering
\tiny
\caption{Decay widths and production cross-sections for a total decay width of $\Gamma = 45$~GeV for some sample points. All masses and decay widths are in GeV. The cross-sections are in femtobarns (fb). The ${\rm Br}_{\rm DM}$ represents the branching ratio allocated to dark matter. For simplicity, we assume here that $M_{XE} = M_{XN}$ and $\lambda_E = \lambda_N$.}
\begin{tabular}{|c|c|c|c|c|c|c|c|c|c|c|c||c|c|c||c|c|c|c||c|c|}\cline{1-21}
\multicolumn{21}{|c|}{$\Gamma = 45~ {\rm GeV}$} \\ \hline
$M_{XU}$&$M_{XD}$&$M_{XE}$&$M_{\widetilde{XU}}$&$M_{\widetilde{XD}}$&$M_{\widetilde{XE}}$&$\lambda_{U}$&$\lambda_{D}$&$\lambda_E$&$A_{U}$&$A_{D}$&$A_E$&$\Gamma_{\gamma \gamma}$&$\Gamma_{gg}$&$\Gamma_{XE+XN}$&${\rm Br}_{\gamma \gamma}$&${\rm Br}_{gg}$&${\rm Br}_{XE+XN}$&${\rm Br}_{\rm DM}$&$\sigma_{\gamma \gamma}^{8~{\rm TeV}}$&$\sigma_{\gamma \gamma}^{13~{\rm TeV}}$\\ \hline \hline
$	855	$&$	735	$&$	265	$&$	1209	$&$	1039	$&$	375	$&$	0.83	$&$	0.83	$&$	0.31	$&$	2565	$&$	2205	$&$	1590	$&$	0.0017	$&$	0.95	$&$	1.02	$&$	0.000037	$&$	0.021	$&$	0.023	$&$	0.956	$&$	0.06	$&$	0.28	$	\\ \hline
$	860	$&$	740	$&$	400	$&$	1216	$&$	1047	$&$	566	$&$	0.83	$&$	0.83	$&$	0.31	$&$	2580	$&$	2220	$&$	2400	$&$	0.0005	$&$	0.93	$&$	0.00	$&$	0.000012	$&$	0.021	$&$	0.000	$&$	0.979	$&$	0.03	$&$	0.14	$	\\ \hline
$	1000	$&$	800	$&$	265	$&$	1414	$&$	1131	$&$	375	$&$	0.83	$&$	0.80	$&$	0.31	$&$	3000	$&$	2400	$&$	1590	$&$	0.0015	$&$	0.71	$&$	1.02	$&$	0.000034	$&$	0.016	$&$	0.023	$&$	0.962	$&$	0.07	$&$	0.31	$	\\ \hline
$	1100	$&$	900	$&$	250	$&$	1556	$&$	1273	$&$	354	$&$	0.83	$&$	0.80	$&$	0.31	$&$	3300	$&$	2700	$&$	1500	$&$	0.0013	$&$	0.56	$&$	1.19	$&$	0.000029	$&$	0.012	$&$	0.026	$&$	0.961	$&$	0.04	$&$	0.21	$	\\ \hline
$	1200	$&$	1100	$&$	265	$&$	1697	$&$	1556	$&$	375	$&$	0.83	$&$	0.83	$&$	0.31	$&$	3600	$&$	3300	$&$	1590	$&$	0.0013	$&$	0.41	$&$	1.02	$&$	0.000029	$&$	0.009	$&$	0.023	$&$	0.968	$&$	0.03	$&$	0.16	$	\\ \hline
\end{tabular}
\label{tab:45GeV}
\end{table}

\noindent where $K$ is the QCD $K$-factor, $\Gamma$ is the total decay width, $\sqrt{s}$ is the proton-proton center of mass energy, and $C_{gg}$ is the dimensionless partonic integral computed for an $M_S = 750$~GeV resonance, yielding $C_{gg} = 174$ at $\sqrt{s} = 8$~TeV and $C_{gg} = 2137$ at $\sqrt{s} = 13$~TeV~\cite{Franceschini:2015kwy}. We use the gluon fusion $K$-factor of $K = 1.98$.

We construct our model with the ($XF$,$\overline{XF}$) and ($Xl$,$\overline{Xl}$) flippons, implementing only one copy of the ($\bf{10}$,$\overline{\bf 10}$). For the calculations, we decompose the $(XQ, XQ^c)$ multiplet into its $(XU, XU^c)$ and $(XD, XD^c)$ components. Given that the flippon multiplets $(XU, XU^c)$, $(XD, XD^c)$, and $(XE, XE^c)$ participate in the $S \rightarrow \gamma \gamma$ loop diagrams and $(XU, XU^c)$, $(XD, XD^c)$ in the $S \rightarrow gg$ loops, there are 12 free parameters in the effective operators $\kappa_{EM}$ and $\kappa_3$ consisting of the Yukawa couplings $\lambda_f$, trilinear A term couplings $A_f$, fermionic component masses $M_f$, and scalar component masses $M_{\widetilde{f}}$. In total, there are 15 parameters to compute:
\begin{equation}
M_{XU},~M_{XD},~M_{XE},~M_{\widetilde{XU}},~M_{\widetilde{XD}},~M_{\widetilde{XE}},~\widetilde{M}_{XU},~\widetilde{M}_{XD},~\widetilde{M}_{XE},~\lambda_U,~\lambda_D,~\lambda_E,~A_U,~A_D,~A_E
\label{list}
\end{equation}
though the supersymmetry breaking soft terms $\widetilde{M}_{XU},~\widetilde{M}_{XD},~\widetilde{M}_{XE}$ can be trivially computed from the fermionic and scalar components using the following relations
\begin{eqnarray}
&& M_{\widetilde{XU}}^2 = M_{XU}^2 + \widetilde{M}_{XU}^2 \\
&& M_{\widetilde{XD}}^2 = M_{XD}^2 + \widetilde{M}_{XD}^2 \\
&& M_{\widetilde{XE}}^2 = M_{XE}^2 + \widetilde{M}_{XE}^2 \
\label{soft_eq}
\end{eqnarray}

The ${\cal F}$-$SU(5)$ flippon parameter space here is rather large, given the freedom on the 12 free-parameters. We equate the fermionic component of the flippon and the flippon soft supersymmetry breaking term, such that $M_f = \widetilde{M}_f$, where consequently the relations above will provide a flippon scalar component a little larger than the fermionic component. The most recent LHC constraints on vector-like $T$ and $B$ quarks~\cite{atlas-vectorlike} establish lower limits of about 855~GeV for $(XQ, ~XQ^c)$ flippons and 735~GeV for $(XD, ~XD^c)$ flippons. Moreover,
the heavy leptons are less efficiently produced at the LHC since they do not necessarily have
significant mixings with SM leptons, and then evade the excited lepton searches~\cite{Khachatryan:2015scf}.
Analogous slepton search bounds are 260 GeV at CMS~\cite{Khachatryan:2014qwa} and 325 GeV at ATLAS~\cite{Aad:2014vma},
with an assumption of large missing transverse energy in the final state, {\it i.e.}, a non-compressed scenario.
Thus, we can take advantage of a light $XE$ multiplet, which can also contribute to invisible branching fractions if $M_{XE} < 375$~GeV. The maximum Yukawa couplings are approximately $\lambda_U \sim \lambda_D \lesssim 0.83$ and  $\lambda_E \lesssim 0.31$, providing an upper limit on their freedom. In order to not break the  $SU(3)_C \times U(1)_{EM}$ gauge symmetry, we limit the $A_{U,D}$ terms to $A_{U,D} \lesssim 3 M_{U,D}$, however, due to the small $XE$ Yukawa coupling $\lambda_E$ we only limit the $A_E$ term to $A_{E} \lesssim 6 M_{E}$. Some sample benchmark points are detailed in TABLE~\ref{tab:5GeV} and TABLE~\ref{tab:45GeV}, providing insight into the coherence amongst the parameters. There are strong constraints on the digluon decay width from $pp \rightarrow jj$ dijets, limiting the digluon decay width to about $\Gamma_{gg} \approx 1.3$~GeV in the model space, though our maximum $A$ terms noted above allow this constraint to be naturally satisfied. Note that the cross-sections $\sigma_{\gamma \gamma}^{8~{\rm TeV}}$ and $\sigma_{\gamma \gamma}^{13~{\rm TeV}}$ in TABLES~\ref{tab:5GeV} - \ref{tab:45GeV} show a gain of 4.65 from 8~TeV to 13~TeV. In our calculations, we take the coupling constants at the scale $M_Z$ to be $\alpha_3 = 0.1185$ and $\alpha_{EM} = 128.91^{-1}$.

The diphoton and digluon modes only comprise a small fraction of the total width, as seen in TABLES~\ref{tab:5GeV} - \ref{tab:45GeV}. These practical limits on the diphoton and digluon decay widths relegate most of the total width to the other decay channels, such as $S \rightarrow XE  XE^c$ and $S \rightarrow XN  XN$ for both $M_{XE}$ and $M_{XN}$ less than 375~GeV and/or to dark matter invisibly. In particular, the decay width for the decay into pairs of fermions
$S \rightarrow f  \overline{f}$ is given by
\begin{equation}
\Gamma(S\rightarrow f \overline{f}) = \frac{1}{16\pi} M_S \lambda_{f}^2 \left(1-\frac{4M_f^2}{M_S^2}\right)^{3/2}.
\end{equation}
As a result, the decay rate for $S \rightarrow XE  XE^c$ is in the range $\Gamma(S\rightarrow XE XE^c) \lesssim 1.2$~GeV, 
and similarly also for the decay rate for $S \rightarrow XN  XN$. Detailed numerical results for these decay channels are listed in TABLES~\ref{tab:5GeV}-\ref{tab:45GeV}, where for simplicity, we assume $M_{XE} = M_{XN}$ and $\lambda_E = \lambda_N$ in the calculations.
Moreover, if we require the total decay width to be 45 GeV, the branching ratio for the $S$ invisible decay into dark matter is close to one and then the monojet searches by the ATLAS and CMS experiments~\cite{Aad:2015zva, Khachatryan:2014rra} have already excluded such a possibility. One solution is the soft lepton approach~\cite{Dutta:2015wqh, Soft-Lepton}. To be concrete, let us present an example
from Ref.~\cite{Soft-Lepton}. We introduce a $Z_2$ symmetry and two SM singlets $N'_k$ with $k=1,~2$,
and assume that under $Z_2$ symmetry, $N'_k$ and $(XE, XE^c)$ are odd while all the other particles are even.
Thus, $N'_k$ and $(XE, XE^c)$ form a new dark matter sector with the lightest particle being $N_1'$,
and we will thus have two dark matter candidates: $N_1'$ and $\chi_1^0$.
The relevant superpotential terms for soft-lepton approach are
\begin{eqnarray}
W &=& M^{N'}_k  N_k' N_k' + \lambda_k' N_k' N_k' + y'_{ik} E_i^c XE N_k~.~
\end{eqnarray}
For simplicity, we assume that the mass difference between $M^{N'}_2$ and $M^{N'}_1$ is from 14~GeV to 20~GeV
and $\lambda_2' >> \lambda_1'$. Thus, we consider $S$ will decay dominantly into a pair of $N'_2$,
which can be semi-invisible and give us a large decay width. Note that if our LSP mass is larger than
200~GeV, the only decay chain for $N_2'$ is $N'_2 \to XE E_i^c \to N_1' E_i^c (E_j^c)^*$. 
Due to the fact that monojet searches veto isolated leptons $(e,~\mu)$ with small $p_T$
$> 7$ GeV and $ p_T(\tau_h)  > 20$~GeV at the ATLAS and CMS experiments~\cite{Aad:2015zva, Khachatryan:2014rra},
 the leptons in the semi-invisible decays will be vetoed during monojet searches, and therefore evade such bounds.
 On the other hand, to trigger the events for the multi-lepton searches at the LHC, we need at least one lepton
 with $P_T$ larger than $26$~GeV and $20$~GeV for the ATLAS and CMS Collaborations,
 respectively~\cite{Aad:2014hja, Chatrchyan:2012mea}.
 Thus, the constraints from multi-lepton searches at the LHC can be evaded as well~\cite{Soft-Lepton}.

\begin{table}[htp]
\centering
\scriptsize
\caption{Sample General No-Scale ${\cal F}$-$SU(5)$ benchmark points with light flippons that satisfy all experimental constraints imposed by the LHC and other essential experiments. All masses are in GeV. The numerical values given for $Br(b \rightarrow s \gamma)$ are $\times 10^{-4}$,  $Br(B_s^0 \rightarrow \mu^+ \mu^-)$ are $\times 10^{-9}$, $\Delta a_{\mu}$ are $\times 10^{-10}$, spin-independent cross-sections $\sigma_{SI}$ are $\times 10^{-12}$~pb, spin-dependent cross-sections $\sigma_{SD}$ are $\times 10^{-10}$~pb, and proton decay rate $\tau_p$ are in units of $10^{35}$ years. The Higgs boson mass $M_h$ calculated here assumes a minimal coupling with the flippon multiplets.}
\begin{tabular}{|c|c|c|c|c|c||c|c|c|c|c|c|c|c|c|c|c|}\cline{1-17}
\multicolumn{17}{|c|}{${\rm General~No-Scale}~{\cal F}-SU(5)$} \\ \hline
$M_{1/2}$&$m_0$&$A_0$&$M_V$&${\rm tan}\beta$&$m_{\rm top}$&$M_{\chi_1^0}$&$M_{\widetilde{\tau}^{\pm}}$&$M_{\widetilde{g}}$&$M_h$&$\Omega h^2$&$Br(b \rightarrow s \gamma)$&$Br(B_s^0 \rightarrow \mu^+ \mu^-)$&$\Delta a_{\mu}$&$\sigma_{SI}$&$\sigma_{SD}$&$\tau_p$\\ \hline \hline
$	1550	$&$	970	$&$	-1000	$&$	750	$&$	41.50	$&$	172.60	$&$	310	$&$	318	$&$	2005	$&$	125.88	$&$	0.1114	$&$	3.38	$&$	4.4	$&$	1.9	$&$	4	$&$	12	$&$	1.1	$	\\ \hline
$	1750	$&$	150	$&$	-1950	$&$	1100	$&$	16.50	$&$	173.00	$&$	358	$&$	360	$&$	2207	$&$	126.04	$&$	0.1285	$&$	3.52	$&$	3.1	$&$	1.0	$&$	2	$&$	5	$&$	0.9	$	\\ \hline
$	1425	$&$	625	$&$	-1625	$&$	1300	$&$	27.25	$&$	173.13	$&$	289	$&$	295	$&$	1806	$&$	126.16	$&$	0.1257	$&$	3.39	$&$	3.4	$&$	1.8	$&$	3	$&$	11	$&$	1.0	$	\\ \hline
$	1912	$&$	160	$&$	-200	$&$	1622	$&$	24.00	$&$	171.87	$&$	400	$&$	402	$&$	2426	$&$	126.15	$&$	0.1205	$&$	3.53	$&$	3.1	$&$	1.2	$&$	3	$&$	12	$&$	0.9	$	\\ \hline
$	1872	$&$	640	$&$	240	$&$	2144	$&$	37.33	$&$	173.34	$&$	397	$&$	399	$&$	2369	$&$	126.17	$&$	0.1117	$&$	3.50	$&$	3.5	$&$	1.6	$&$	4	$&$	16	$&$	0.9	$	\\ \hline
$	1862	$&$	630	$&$	980	$&$	2966	$&$	40.66	$&$	172.97	$&$	400	$&$	402	$&$	2357	$&$	124.83	$&$	0.1112	$&$	3.50	$&$	3.7	$&$	1.9	$&$	7	$&$	26	$&$	0.9	$	\\ \hline\end{tabular}
\label{tab:GNS}
\end{table}

An interesting inquiry here is whether such a model with light flippons is realistic with a viable SUSY spectrum at the LHC II, in addition to satisfying all other complementary constraints from other essential experiments running in parallel with the LHC. These encompass LHC gluino and light Higgs boson mass constraints, direct detection of dark matter, relic density measurements, proton decay rate $\tau_p$ via $p \rightarrow e^+ \pi^0$, and rare-decay processes ($b \rightarrow s \gamma$, $B_s^0 \rightarrow \mu^+ \mu^-$, $\Delta a_{\mu}$). When incorporated into the General No-Scale ${\cal F}$-$SU(5)$ GUT~\cite{HLLMN-P}, a light vector-like mass is indeed viable, as exemplified by the benchmarks scenarios listed in TABLE~\ref{tab:GNS} that employ CMSSM boundary conditions with a flippon mass $M_V$~\cite{HLLMN-P}. Note though that the $(XU, XU^c)$, $(XD, XD^c)$, and $(XE, XE^c)$ flippon multiplets in TABLE~\ref{tab:GNS} computed within the General No-Scale ${\cal F}$-$SU(5)$ GUT are assumed to have a universal mass $M_V$, with no mass splitting as applied in TABLE~\ref{tab:5GeV} and TABLE~\ref{tab:45GeV}. The renormalization group running of flippon multiplets with mass splitting will be the goal of a future project, though the introductory results given here in TABLE~\ref{tab:GNS} indicate that the SUSY spectrum and experimental constraints are anticipated to remain viable regardless of whether the flippon masses are universal or split.

\section{Conclusions}

We presented here an explanation for the diphoton excesses recently observed at the $\sqrt{s} = 13$~TeV LHC II indicating a possible 750~GeV resonance. Our methodology engages the realistic supersymmetric GUT model ${\cal F}$-$SU(5)$ with additional vector-like multiplets, denoted flippons. Though the statistical significance of the data bump remains inconclusive, it is nonetheless compelling given the correlation between excess diphoton events observed by both the ATLAS and CMS Collaborations, and also the consistency between the 8~TeV and 13~TeV data at two standard deviations. It is thus a worthwhile endeavor to build a realistic model capable of interpreting a 750~GeV resonance within a consistent GUT.

Our primary ingredient are the additional vector-like particles, which we refer to as flippons. Flippons were shown previously in the No-Scale ${\cal F}$-$SU(5)$ GUT to provide rather favorable phenomenology at the LHC and all experiments, and here they perform another phenomenological feat. In the ${\cal F}$-$SU(5)$ framework, a 750~GeV singlet field produced by a gluon fusion triangle diagram can then decay to a diphoton via flippon loops. This is all accomplished with realistic electric charges and Yukawa couplings, given the ${\cal F}$-$SU(5)$ GUT is presently being probed at the LHC. The numerical analysis presented showed that light flippon masses in the loops can produce the observed cross-sections. We showed previously in the $one$-$parameter$ No-Scale ${\cal F}$-$SU(5)$ ($m_0 = A_0 = B = 0$) that heavy flippons are required to maintain experimental viability. Here in this work we gave some sample benchmark scenarios in General No-Scale ${\cal F}$-$SU(5)$ offering evidence that indeed light flippons with their associated SUSY spectrum can also be experimentally viable. Given this favorable phenomenology, we believe the work demonstrated here represents a realistic explanation of the diphoton excesses at an invariant mass of about 750~GeV.


\begin{acknowledgments}

This research was supported in part by the Natural Science Foundation of China under grant numbers 11135003, 11275246, and 11475238 (TL), and by the DOE grant DE-FG02-13ER42020 (DVN).

\end{acknowledgments}

\end{document}